\documentclass[showpacs,prd,eqsecnum,preprintnumbers,amsmath,nofootinbib,amssymb,eqsecnum]{revtex4}
\usepackage{amsmath,amsthm,amssymb}
\usepackage{graphics}
\usepackage{graphicx}
\usepackage{subfigure}
\usepackage{dcolumn}
\begin{document}
\preprint{}
\title{A trapped surface in the higher-dimensional self-similar Vaidya spacetime}
\author{${}^{1,2}$Masahiro Shimano}
 \email{mshimano@rikkyo.ac.jp}
\author{${}^{2}$Tomohiro Harada}
  \email{harada@rikkyo.ac.jp}
\author{${}^{2}$Naoki Tsukamoto}
  \email{11ra001t@rikkyo.ac.jp}
\affiliation{
${}^{1}$Jumonji Junior and Senior High School, Toshima, Tokyo 170-0004, Japan
\\${}^{2}$Department of Physics, Rikkyo University, Toshima, Tokyo 171-8501, Japan}
\date{\today}
\begin{abstract}

We investigate a trapped surface and naked singularity in 
a $D$-dimensional Vaidya spacetime with a self-similar mass function.
A trapped surface is defined as 
a closed spacelike $(D-2)$-surface which has negative both null expansions.
There is no trapped surface in the Minkowski spacetime.
However, in a four-dimensional self-similar Vaidya spacetime, 
Bengtsson and Senovilla considered non-spherical trapped surfaces 
and showed that a trapped surface can penetrate into a flat region, 
if and only if the mass function rises fast enough~[I. Bengtsson and J. M. M. Senovilla, Phys. Rev. D \textbf{79}, 024027 (2009).].
We apply this result to a $D$-dimensional spacetime 
motivated by 
the context of large extra dimensions or TeV-scale gravity.
In this paper, similarly to Bengtsson and Senovilla's study, 
we match four types of $(D-2)$-surfaces and show that
a trapped surface extended into the flat region can be 
constructed 
in the $D$-dimensional Vaidya spacetime,
if the increasing rate of the mass function is greater than 
$0.4628$.
Moreover, we show that
the maximum radius of the trapped surface constructed here
approaches the Schwarzschild-Tangherlini radius in the large $D$ limit.
Also, we show that 
there is no naked singularity, 
if the spacetime has the trapped surface constructed here.

\end{abstract}

\maketitle

\section{Introduction}

The boundary of a region in a spacetime
that cannot be observed from infinity is called an event horizon.
The event horizon is the region of the boundary of a black hole
which has a teleological property:
the entire future history of the spacetime must be known before the position 
of the event horizon can be determined.
Black holes might be formed by some dynamical process, and 
then might undergo accretions and evolutionary processes.
In numerical relativity, 
to investigate the evolution of the black hole 
we need to identify the boundary of the black hole 
in an initial data set.
From this context
it is difficult to investigate the evolution of 
the black hole by using the event horizon
in numerical relativity.

Eardley conjectured that the boundary of the region which contains marginally outer trapped 
surfaces coincides with the event horizon~\cite{D.M.Eardley}, where 
an outer trapped surface 
is defined as a closed spacelike two-surface~(in the four-dimensional case) 
which has a negative outer null expansion. 
Although the event horizon has the teleological notion
and is defined in terms of future null infinity, 
Eardley's conjecture suggests that 
the event horizon can be constructed by the outer 
trapped surface without the teleological notion.
In a four-dimensional Vaidya spacetime, 
Ben-Dov showed that Eardley's conjecture is true~\cite{I.Ben-Dov}. 
To study this conjecture in various spacetimes might be important to 
understand dynamical black holes.
However, the outer trapped surface cannot be defined in general spacetimes, 
while it is defined only in asymptotically flat spacetimes~\cite{Hawking}.
Moreover, the outer trapped surface is only defined by outgoing null rays.
We do not know whether the outer trapped surface has a negative ingoing 
null expansion or not. 
To resolve this difficulty we consider a trapped surface, where
a trapped surface is defined as the closed spacelike two-surface 
which has negative both null expansions.
In general spacetimes, while ingoing and outgoing null rays 
are not defined, 
there are two independent null rays.
Both null expansions are defined by these two independent null rays.
Thus, the trapped surface can be defined not only 
in asymptotically flat spacetimes but also in various spacetimes.

As is well known, there is no trapped surface in the Minkowski spacetime. 
However, in the four-dimensional Vaidya spacetime 
it was shown that a trapped surface can 
penetrate into a flat region.
The numerical results of Schnetter and Krishnan showed that the outer boundary of 
trapped surfaces can penetrate into the flat region~\cite{E. Schnetter and B. Krishnan}.
Moreover, Bengtsson and Senovilla considered the self-similar Vaidya 
spacetime, and they analytically showed that a trapped surface can 
penetrate into the flat region, 
if and only if the mass function rises fast enough~\cite{I.Bengtsson}.
Is this feature common in higher-dimensional spacetimes?
In the present paper, we focus on a higher-dimensional Vaidya spacetime
and investigate a trapped surface in this spacetime.

Recently, higher-dimensional scenarios with large~\cite{ArkaniHamed:1998rs} 
and warped~\cite{Randall:1999ee} extra dimensions were proposed to resolve the hierarchy problem between 
the gravitational and electroweak interactions.
One of the most striking predictions of such scenarios is the production of the large number of 
mini black holes in high-energy particle collisions~\cite{Argyres:1998qn}.
Therefore, to study higher-dimensional black holes is important 
in the context of above scenarios.

Can we construct a trapped surface extended 
into the flat region in the higher-dimensional spacetime as 
in the four-dimensional spacetime?
In the present paper, we apply Bengtsson and Senovilla's study to an 
$(n+3)$-dimensional ($n=D-3\geq1$) Vaidya spacetime with a self-similar mass function. 
We concern a trapped surface constructed by matching four types
of $(n+1)$-surfaces and show that a trapped surface can penetrate into the flat region. 
Moreover, we demonstrate that there is no naked singularity, 
if the spacetime has the trapped surface constructed here.

This paper is organized as follows.
In Sec.~\ref{sec:vaidya}, we briefly review the $(n+3)$-dimensional self-similar Vaidya spacetime
and mention the condition for the black hole and naked singularity in this spacetime.
In Sec.~\ref{sec:expansion}, we introduce two classes of $(n+1)$-surfaces to construct 
a trapped surface, and calculate both null expansions of these surfaces.
In Sec.~\ref{sec:trapped}, we consider four types of $(n+1)$-surfaces and match these surfaces. 
Then, we construct a trapped surface extended into the flat region and 
show the condition to construct this surface.
Moreover, we discuss the relation of the condition 
for the trapped surface and naked singularity, 
and demonstrate that there is no naked singularity
if the spacetime has the trapped surface constructed here.
We conclude the paper in Sec.~\ref{sec:conclusion}.

\section{$(n+3)$-dimensional Vaidya spacetime}\label{sec:vaidya}

We focus on the $(n+3)$-dimensional ($n=D-3\geq 1$) Vaidya spacetime~\cite{B.R.Iyer and C.V.Vishveshwara}
	\begin{equation}\label{eq:metric}
	 ds^2=-\left(1-\frac{2m(v)}{nr^{n}}\right)dv^2+2dvdr+r^2d\Omega^2_{n+1},
	\end{equation}
where $m(v)$ is the mass function, and 
	\begin{equation}
	 d\Omega^2_{n+1}=d\theta_1^2+\sin^2\theta_1d\theta_2^2+\cdots +\left(\prod^n_{i=1}\sin^2\theta_i\right)d\varphi^2 
	\end{equation}
is the line element on an unit $(n+1)$-sphere.
$\theta_i$ is an inclination angle defined in $0\leq \theta_i\leq \pi$ 
and $\varphi$ is an azimuthal angle defined in $0\leq \varphi\leq 2\pi$, respectively.
For an arbitrary $m(v)$, the following stress energy tensor solves the 
Einstein equation,
	\begin{equation}
         T_{\mu\nu}=\frac{(n+1)\dot{m}}{8\pi}l_\mu l_\nu,
	\end{equation}
where $l_\mu=-\nabla_\mu v$ is an ingoing null vector, 
$\nabla_\mu$ is the covariant derivative in the $(n+3)$-dimensional spacetime, 
and the dot is the differentiation with respect to $v$.
In the present paper,
we choose the mass function such as
	\begin{equation}
	 m=\Bigg\{
         \begin{array}{ccl}
         0&\mathrm{for}&\quad v\leq 0\\
         \mu v^{n}&\mathrm{for} &\quad 0\leq v\leq v_0\\
         M&\mathrm{for} &\quad v\geq v_0
         \end{array},
	\end{equation}
where $v_0=M^{1/n}/\mu$, 
$\mu$ and $M$ are positive constants, respectively.
We call $\mu$ {\it a mass parameter}.
There is the radial influx of a null fluid for an initially empty region. 
We call the region $v<0$ {\it region I}.
The region $0\leq v\leq v_0$ has the radial influx.
We call this region {\it region II}.
The region $v\geq v_0$ has a constant mass is called {\it region III}.
These regions are shown in Fig.~1.
Note that 
naked singularity occurs if and only if
the mass parameter satisfies the following condition~\cite{S.G.Ghosh and N.Dadhich}:
	\begin{equation}\label{def:ns}
	 0<\mu\leq \mu_c=\left\{\frac{n}{2(n+1)}\right\}^{n+1}.
	\end{equation}
For $\mu>\mu_c$, singularity is not even naked, i.e., 
the spacetime has the black hole. 
Conformal diagrams of these spacetimes are drawn in Fig.~1.

\begin{figure}[h!]
 \begin{center}
  \subfigure[]{
   \includegraphics[width=.42 \columnwidth]{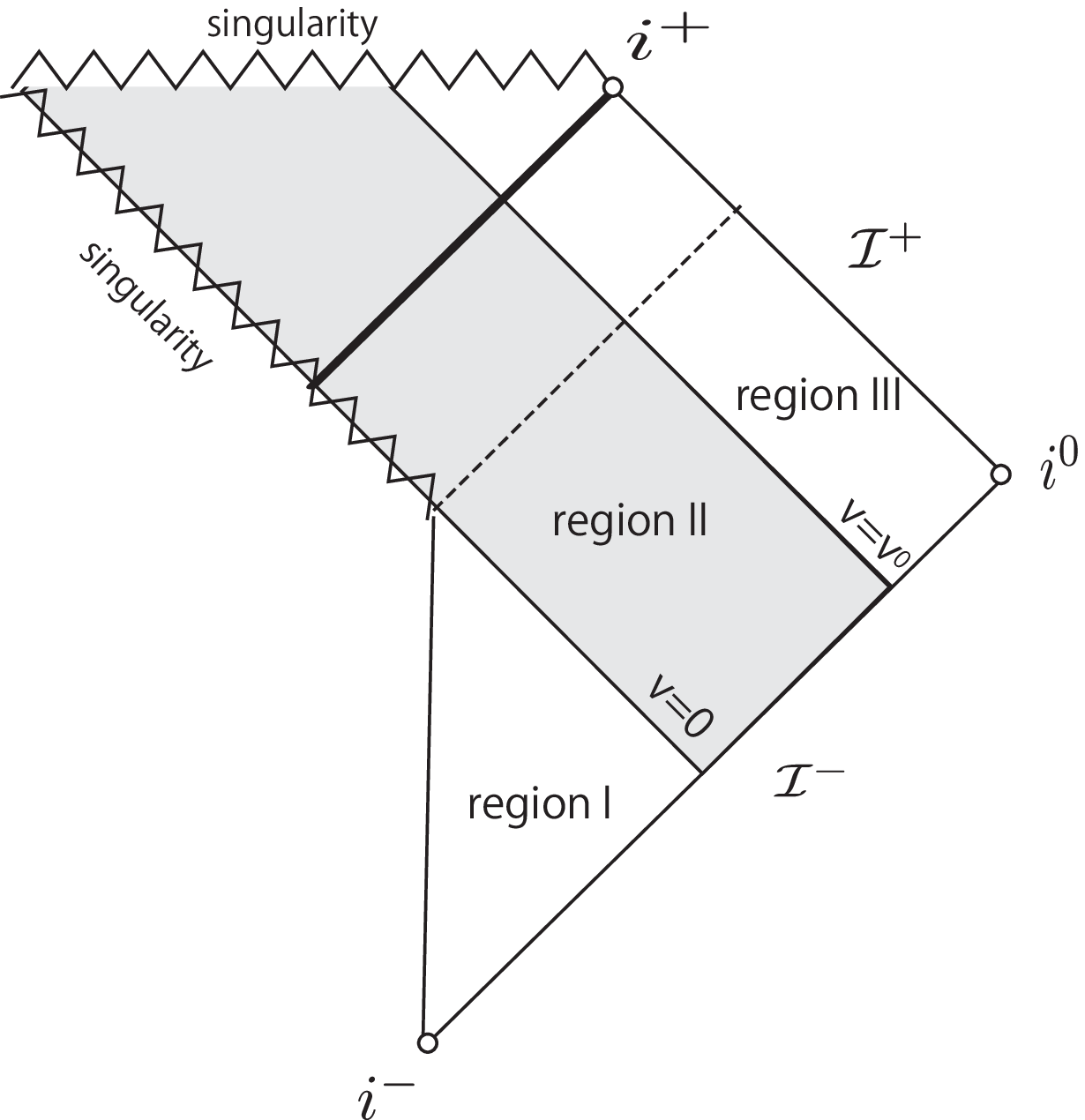}
  \label{fig:pd}}~\hspace{3cm}
  \subfigure[]{
   \includegraphics[width=.32 \columnwidth]{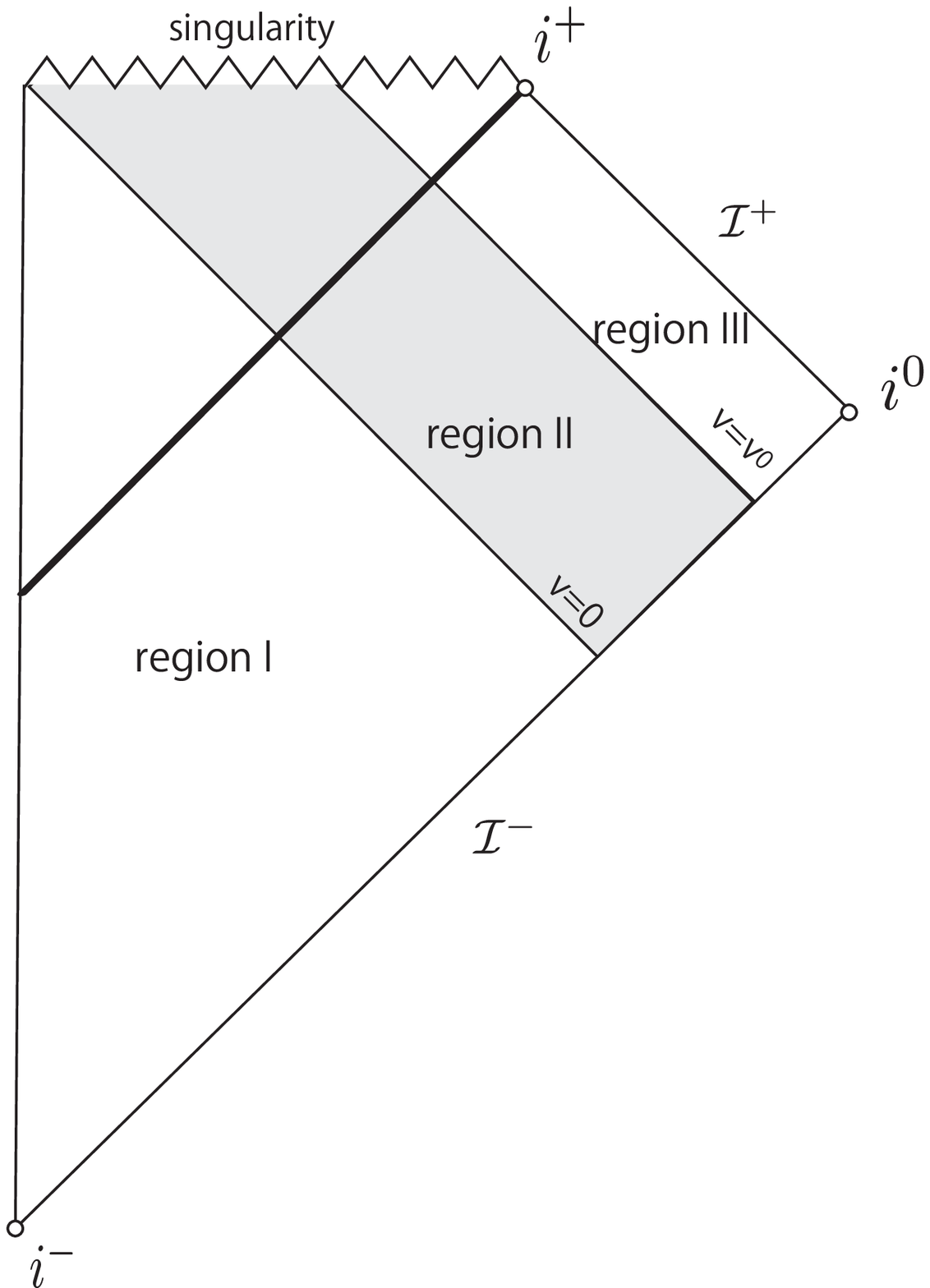}
  \label{fig:pd2}}
  \caption{
  Conformal diagrams of the $(n+3)$-dimensional Vaidya spacetime.
  The shaded region is region II.
  This region is bounded by region I on $v=0$ and region III on $v=v_0$.
  (a) The spacetime collapses to globally naked singularity, where the mass parameter 
  satisfies $0<\mu\leq\mu_c$.
  The event horizon and the Cauchy horizon are shown as the bold and the dashed lines.
  (b) The spacetime collapses to the black hole, where the mass parameter 
  satisfies $\mu>\mu_c$.
  The thick solid line is the event horizon.
        } 
  \label{fig:1}
 \end{center}
\end{figure}

\section{Both null expansions for two classes of $(n+1)$-surfaces}\label{sec:expansion}

In Bengtsson and Senovilla's study
they concerned the four-dimensional self-similar Vaidya spacetime and 
used two classes of two-surfaces to construct a trapped surface 
extended into the flat region.
To apply this study to the case of the $(n+3)$-dimensional spacetime
we introduce two classes of $(n+1)$-surfaces as in Table~1.
\begin{table}[h!]
\caption{The class of $(n+1)$-surfaces where $i,j = 1, \ldots, n$}
\begin{center}
\begin{tabular}{| c | c | c | c | c | c |}
 \hline  class & 
 $ v$ & 
 $ r$ & 
 $ \theta_i$ &  
 $\theta_{j\neq i}$&
 $\varphi$\\ 
 \hline A& $ V(\rho)$ & $R(\rho)$ & $ \pi/2$ & $0\leq\theta_{j}\leq\pi$& $0\leq\varphi\leq 2\pi$ \\
 \hline B& $ V(\rho)$ & $r_0$ & $\Theta_{i}(\rho)$& $0\leq\theta_{j}\leq\pi$& $0\leq\varphi\leq 2\pi$ \\ \hline
\end{tabular}
\end{center}
\end{table}
We call the surface in which 
$r$ and $v$ are the function of $\rho$, 
and $\theta_i=\pi/2$ 
{\it class A}, 
where $i=1,\ldots, n$.
Similarly, 
we call the surface in which 
$\theta_i$ and $v$ are the function of $\rho$, 
and $r=r_0$ {\it class B}, where $i=1,\ldots, n$. 
We shall calculate both null expansions for these two classes of $(n+1)$-surfaces.

\subsection{Both null expansions for a class A surface}\label{sec:expansion_CIS}

We introduce the following the $(n+1)$-surface of class A:
	\begin{equation}\label{case1}
	 \theta_1=\frac{\pi}{2},\hspace{5mm}r=R(\rho),\hspace{5mm}v=V(\rho),
         \hspace{5mm}\theta_2=\phi_2,\quad\cdots,\quad\theta_{n}=\phi_{n},
         \hspace{5mm}\varphi=\phi_{n+1},
	\end{equation}
where we have chosen $i=1$.
The first fundamental form on this surface is 
	\begin{equation}
	 d\gamma^2=\Delta d\rho^2+R^2d\Omega_{n}^2,
	\end{equation}
where $d\Omega^2_{n}$ is the line element on an unit $n$-sphere, 
we have put 
	\begin{equation}
	 \Delta=V'\left(2R'+\Sigma V'\right),
	\end{equation}
and
	\begin{equation}\label{eq:def_sigma}
	 \Sigma=\frac{2m}{nR^{n}}-1,
	\end{equation}
and the prime denotes the differentiation with respect to $\rho$. 
To obtain a spacelike $(n+1)$-surface we demand $\Delta>0$, 
and hence we must have $V'\neq 0$.
We choose orthonormal basis vectors tangent to this $(n+1)$-surface as follows:
	\begin{align}\label{tangent_vector01}
	 &\nonumber Y_{(1)}^\mu=\frac{V'}{\sqrt{\Delta}}\left(\frac{\partial}{\partial v}\right)^\mu
         +\frac{R'}{\sqrt{\Delta}}\left(\frac{\partial}{\partial r}\right)^\mu,\quad
         Y_{(2)}^\mu=\frac{1}{R}\left(\frac{\partial}{\partial \theta_2}\right)^\mu,\\
	 &Y_{(J)}^\mu= \frac{1}{R}\left(\prod_{k=2}^{J-1}\frac{1}{\sin\phi_k}\right)\left(\frac{\partial}{\partial \theta_{J}}\right)^\mu,\quad
         Y_{(n+1)}^\mu= \frac{1}{R}\left(\prod_{k=2}^{n}\frac{1}{\sin\phi_k}\right)\left(\frac{\partial}{\partial \varphi}\right)^\mu,
	\end{align}
where $J=3,\ldots, n$.
Also, we choose normal vectors of this $(n+1)$-surface as 
	\begin{equation}
         n_\mu=\frac{1}{\sqrt{\Delta}}\left\{-R'(dv)_\mu+V'(dr)_\mu\right\},\quad e_\mu=R(d\theta_1)_\mu, 
        \end{equation}
where $n_\mu$ and $e_\mu$ are timelike and spacelike vectors, respectively.
These vectors satisfy conditions $-n^2=e^2=1$ and $n\cdot e=0$.
Using these normal vectors, we obtain null normals as
	\begin{equation}\label{null_normal01}
         N^{(+)}_\mu=\frac{1}{\sqrt{2}}\left(n_\mu+e_\mu\right),\quad N^{(-)}_\mu=\frac{1}{\sqrt{2}}\left(n_\mu-e_\mu\right).
        \end{equation}
Both null expansions are given by~\cite{Hawking}
	\begin{equation}\label{both_null_expansion}
         \vartheta_{(\pm)}=\left(\nabla_\mu N^{(\pm)}_{\nu}\right)\sum_{I=1}^{n+1}Y_{(I)}^{\mu}Y_{(I)}^{\nu}.
        \end{equation}
Substituting orthonormal basis vectors~(\ref{tangent_vector01}) and null normals~(\ref{null_normal01}) into
Eq.~(\ref{both_null_expansion}), and after some calculation, 
we obtain both null expansions as follows:
	\begin{equation}\label{eq:type1_0}
	 \vartheta_{(\pm)}=\frac{1}{\sqrt{2\Delta}}\left\{\frac{1}{\Delta}
         \left(R'V''-V'R''+\frac{mV'^2R'}{R^{n+1}}
         -\frac{\dot{m}V'^3}{nR^{n}}\right)-\frac{nR'}{R}
         +\frac{V'}{R}\left(n-\frac{m}{R^{n}}\right)\right\}.
	\end{equation}	
Note that 
we can introduce class A surfaces for $i\neq1$ and 
can calculate these both null expansions. 
Then, we find that the both null expansions for these surfaces are 
written in the same form as Eq.~(\ref{eq:type1_0}).

\subsection{Both null expansions for a class B surface}\label{sec:uniform_rad}

We introduce the $(n+1)$-surface of class B 
 	\begin{equation}\label{case2}
	 \theta_1=\Theta_{1}(\rho),\hspace{5mm}v=V(\rho),
         \hspace{5mm}r=r_0,\hspace{5mm}\theta_2=\phi_2,\quad\cdots,\quad\theta_{n}=\phi_{n},\quad\varphi=\phi_{n+1},
	\end{equation}
where $r_0$ is a positive constant, and 
we have chosen $i=1$.
The first fundamental form is 
	\begin{equation}
	 d\gamma_{1}^2=\Delta_{1} d\rho^2+r_0^2\sin^2\Theta_{1} d\Omega_{n}^2,
	\end{equation}
where we have put 
	\begin{equation}
	 \Delta_{1}=\Sigma V'^2+r_0^2\Theta_{1}'^2,
	\end{equation}
and $\Sigma$ is written in the same form as Eq.~(\ref{eq:def_sigma}).
To obtain the spacelike $(n+1)$-surface we also demand $\Delta_{1}>0$. 
We choose orthonormal basis vectors tangent to this surface as follows:
	\begin{align}\label{tangent_vector02}
         &\nonumber{Y}_{(1)}^\mu=\frac{V'}{\sqrt{\Delta_{1}}}\left(\frac{\partial}{\partial v}\right)^\mu
         +\frac{\Theta_{1}'}{\sqrt{\Delta_{1}}}\left(\frac{\partial}{\partial \theta_1}\right)^\mu, \quad
         {Y}_{(2)}^\mu=\frac{1}{r_0\sin\Theta_{1}}\left(\frac{\partial}{\partial \theta_2}\right)^\mu,\\
         &{Y}_{(J)}^\mu=\frac{1}{r_0\sin\Theta_{1}}\left(\prod_{k=2}^{J-1}\frac{1}{\sin\phi_k}\right)
         \left(\frac{\partial}{\partial \theta_{J}}\right)^\mu,\quad
         {Y}_{(n+1)}^\mu=\frac{1}{r_0\sin\Theta_{1}}\left(\prod_{k=2}^{n}\frac{1}{\sin\phi_k}\right)
         \left(\frac{\partial}{\partial \varphi}\right)^\mu,
	\end{align}
where $J=3, \ldots, n$.
On the other hand, 
null normals of this surface are  
        \begin{align}\label{null_normal02}
         &\nonumber{N}^{(+)}_\mu=-\frac{r_0\Theta_{1}'\sqrt{\Sigma}}{\sqrt{2\Delta_{1}}}(dv)_\mu
         +\frac{1}{\sqrt{2\Sigma}}\left(1-\frac{r_0\Theta_{1}'}{\sqrt{\Delta_{1}}}\right)(dr)_\mu
         +\frac{r_0V'\sqrt{\Sigma}}{\sqrt{2\Delta_{1}}}(d\theta_1)_\mu,\\
         &{N}^{(-)}_\mu=\frac{r_0\Theta_{1}'\sqrt{\Sigma}}{\sqrt{2\Delta_{1}}}(dv)_\mu
         +\frac{1}{\sqrt{2\Sigma}}\left(1+\frac{r_0\Theta_{1}'}{\sqrt{\Delta_{1}}}\right)(dr)_\mu
         -\frac{r_0V'\sqrt{\Sigma}}{\sqrt{2\Delta_{1}}}(d\theta_1)_\mu.
	\end{align}
Substituting Eqs.~(\ref{tangent_vector02}) and (\ref{null_normal02}) 
into Eq.~(\ref{both_null_expansion}),
we obtain both null expansions of a class B surface for $i=1$ as follows:
	\begin{eqnarray}\label{typeB1}
	 \vartheta_{1}{}_{(\pm)}{}&=&\frac{\sqrt{\Sigma}}{\sqrt{2\Delta_{1}}\Delta_{1}}\left[\pm r_0\left(\Theta_{1}'V''-V'\Theta_{1}''\right)
         +\frac{\sqrt{\Delta_{1}}}{r_0}\left\{n\left(1-\frac{m}{nr_0^{n}}\right)V'^2-(n+1)r_0^2\Theta_{1}'^2\right\}\right.\\
         \nonumber &&\left.\pm n\frac{\Delta_{1} V'}{r_0}\cot\Theta_{1}\right]-\frac{\dot{m}V'^2\left(\sqrt{\Delta_{1}}\mp r_0\Theta_{1}'\right)}
         {\sqrt{2\Sigma\Delta_{1}}\Delta_{1}nr_0^{n}}.
	\end{eqnarray}
Also, from a similar calculation 
both null expansions of the class B surface for $i\neq 1$ can be obtained as follows:
	\begin{eqnarray}\label{typeBall}
	 \nonumber \vartheta_{i}{}_{(\pm)}&=&\frac{\sqrt{\Sigma}}{\sqrt{2\Delta_{i}}\Delta_{i}}\Big[\pm r_0\left(\Theta_{i}'V''-V'\Theta_{i}''\right)
         \left(\prod_{k=1}^{i-1}\sin\phi_k\right)
         +\frac{\sqrt{\Delta_{i}}}{r_0}\left\{n\left(1-\frac{m}{nr_0^{n}}\right)V'^2
         -(n+1)r_0^2\Theta_{i}'^2\left(\prod_{k=1}^{i-1}\sin^2\phi_k\right)\right\}\\
         &&\pm n\frac{\Delta_{i} V'}{r_0}\cot\Theta_{i}\left(\prod_{k=1}^{i-1}\frac{1}{\sin\phi_k}\right)\Big]-\frac{\dot{m}V'^2\left(\sqrt{\Delta_{i}}\mp r_0\Theta_{i}'\right)}
         {\sqrt{2\Sigma\Delta_{i}}\Delta_{i}nr_0^{n}}, 
	\end{eqnarray}
where $i=2,\ldots, n$, $\theta_{i}=\Theta_i(\rho)$, 
and we have put 
	\begin{equation}
	 \Delta_{i}=\Sigma V'^2+r_0^2\Theta_{i}'^2\prod_{k=1}^{i-1}\sin^2{\phi_k}.
	\end{equation}
We also have demanded $\Delta_i>0$ to obtain the spacelike surface.
\section{construction of a trapped surface}\label{sec:trapped}

In this section, 
using surfaces of class A and B, 
we present four types of $(n+1)$-surfaces and 
construct a trapped surface extended into region I. 
In region I, we introduce the $(n+1)$-surface 
which is one of class A surfaces and is a 
topological disk given by the hyperboloid. 
We call this surface {\it type A1}.
In region~II, we introduce the $(n+1)$-surface which is also of class A 
and call this surface {\it type A2}.
In region~III, we consider two types of $(n+1)$-surfaces.
The first one has the property of classes A and B, and we call this surface {\it type AB}.
The second is class B and is called {\it type B1}.
These types of $(n+1)$-surfaces are shown in Fig.~2.
We find the condition for each type of $(n+1)$-surfaces 
to have negative both null expansions.
We also find the condition for these four types of $(n+1)$-surfaces 
to consist a smooth closed $(n+1)$-surface. 
We take an intersection of these conditions
and discuss naked singularity.

\begin{figure}[ht]\label{fig:type}
 \begin{tabular}{c|cccc}
 &
 \begin{minipage}{0.1\hsize}
  \begin{center}
   \includegraphics[width=0.15in,clip]{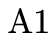}
  \end{center}
 \end{minipage} &
 \begin{minipage}{0.1\hsize}
  \begin{center}
   \includegraphics[width=0.15in,clip]{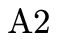}
  \end{center}
 \end{minipage} &
 \begin{minipage}{0.2\hsize}
  \begin{center}
   \includegraphics[width=0.15in,clip]{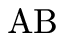}
  \end{center}
 \end{minipage} &
 \begin{minipage}{0.2\hsize}
  \begin{center}
   \includegraphics[width=0.15in,clip]{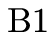}
  \end{center}
 \end{minipage} \\
\hline
 \\
 \begin{minipage}{0.06\hsize}
  \begin{center}
   \includegraphics[width=0.4in,clip]{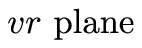}
  \end{center}
 \end{minipage} &
 \begin{minipage}{0.2\hsize}
  \begin{center}
   \includegraphics[width=1in,clip]{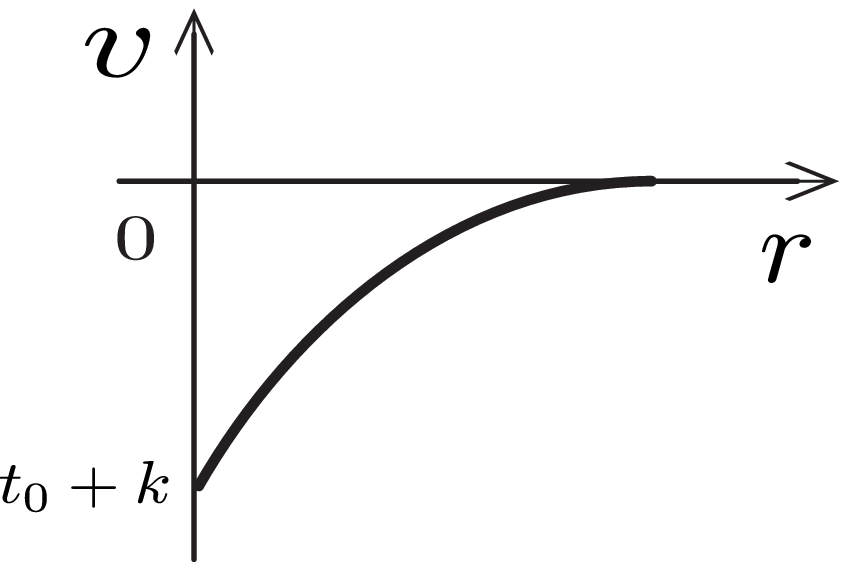}
  \end{center}
 \end{minipage} &
 \begin{minipage}{0.2\hsize}
  \begin{center}
   \includegraphics[width=1in,clip]{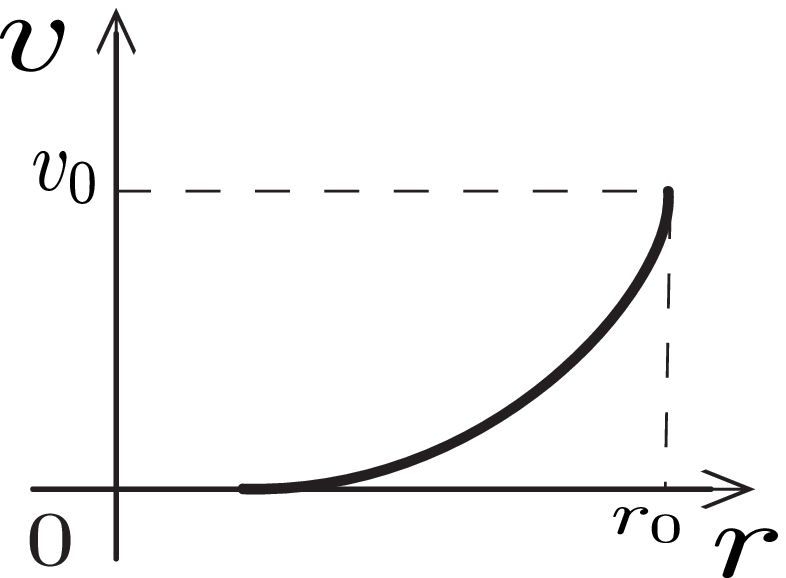}
  \end{center}
 \end{minipage} &
 \begin{minipage}{0.2\hsize}
  \begin{center}
   \includegraphics[width=1in,clip]{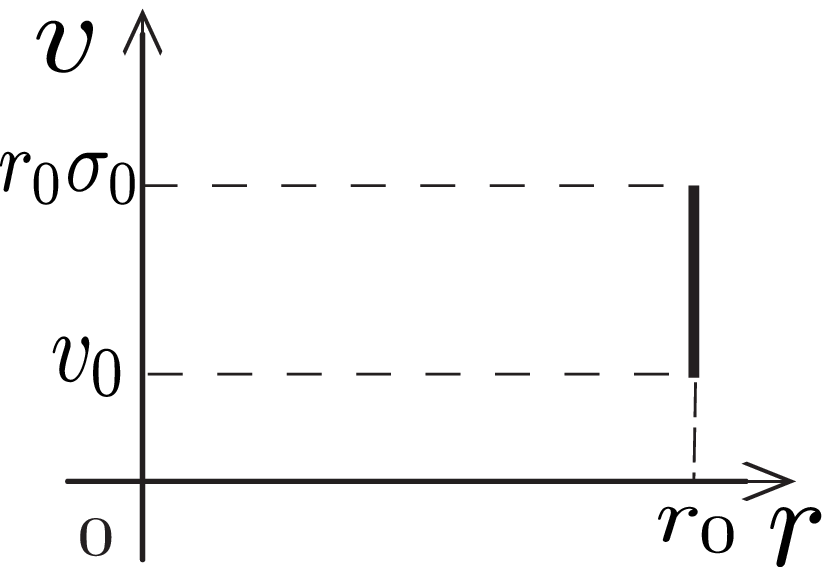}
  \end{center}
 \end{minipage} &
 \begin{minipage}{0.2\hsize}
  \begin{center}
   \includegraphics[width=1in,clip]{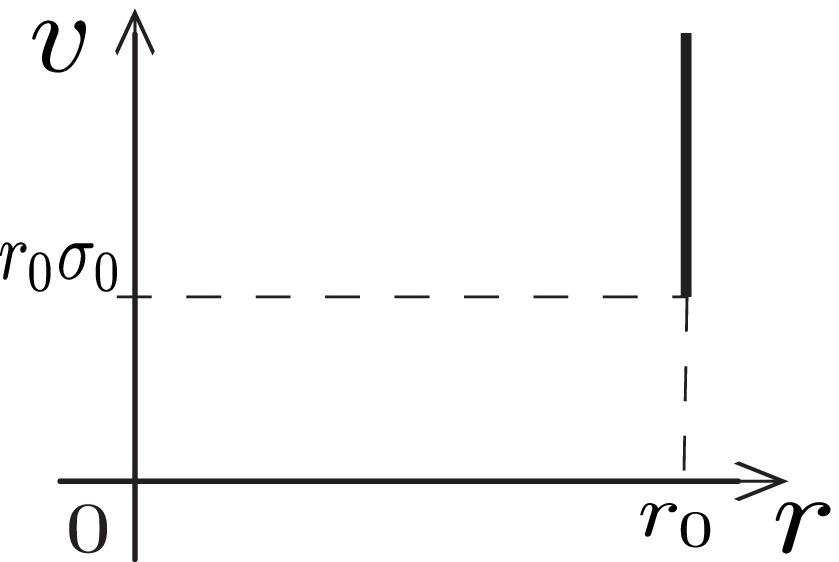}
  \end{center}
 \end{minipage} \\
 \\
 \begin{minipage}{0.06\hsize}
  \begin{center}
   \includegraphics[width=0.4in,clip]{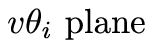}
  \end{center}
 \end{minipage} &
 \begin{minipage}{0.2\hsize}
  \begin{center}
   \includegraphics[width=1in,clip]{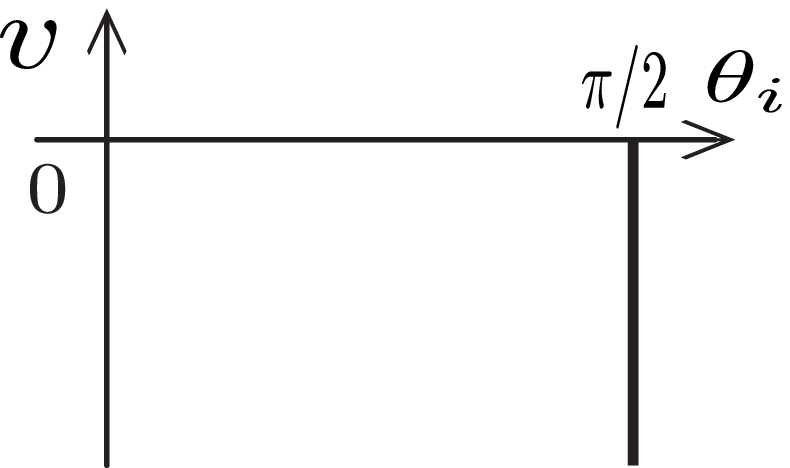}
  \end{center}
 \end{minipage} &
 \begin{minipage}{0.2\hsize}
  \begin{center}
   \includegraphics[width=1in,clip]{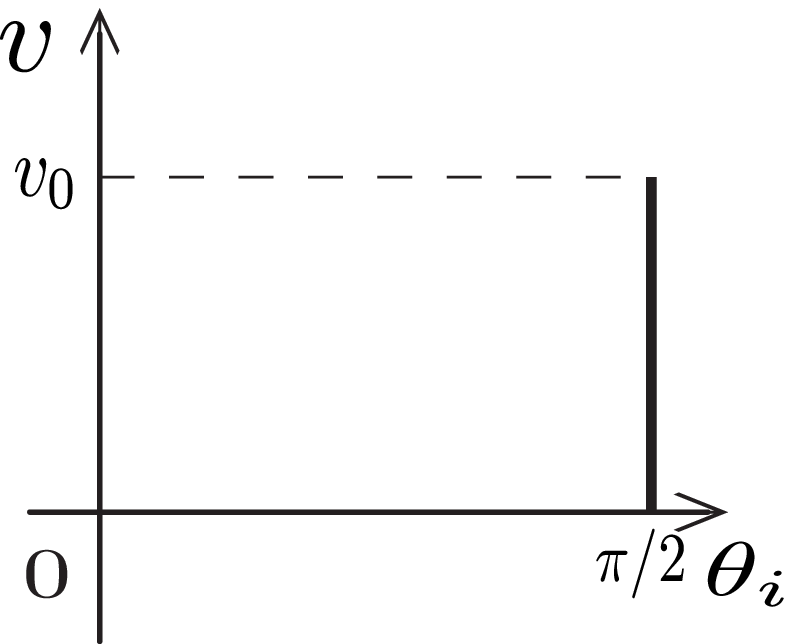}
  \end{center}
 \end{minipage} &
 \begin{minipage}{0.2\hsize}
  \begin{center}
   \includegraphics[width=1in,clip]{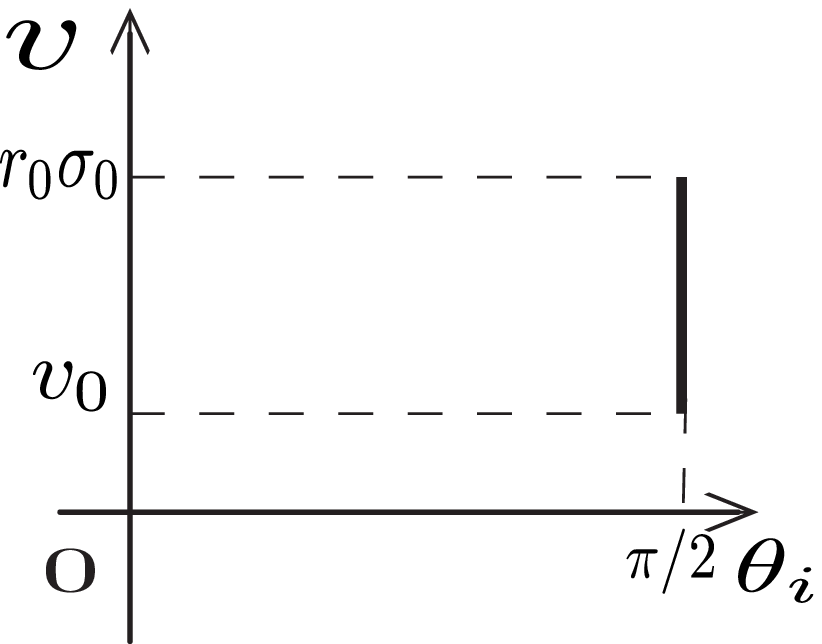}
  \end{center}
 \end{minipage} &
 \begin{minipage}{0.2\hsize}
  \begin{center}
   \includegraphics[width=1in,clip]{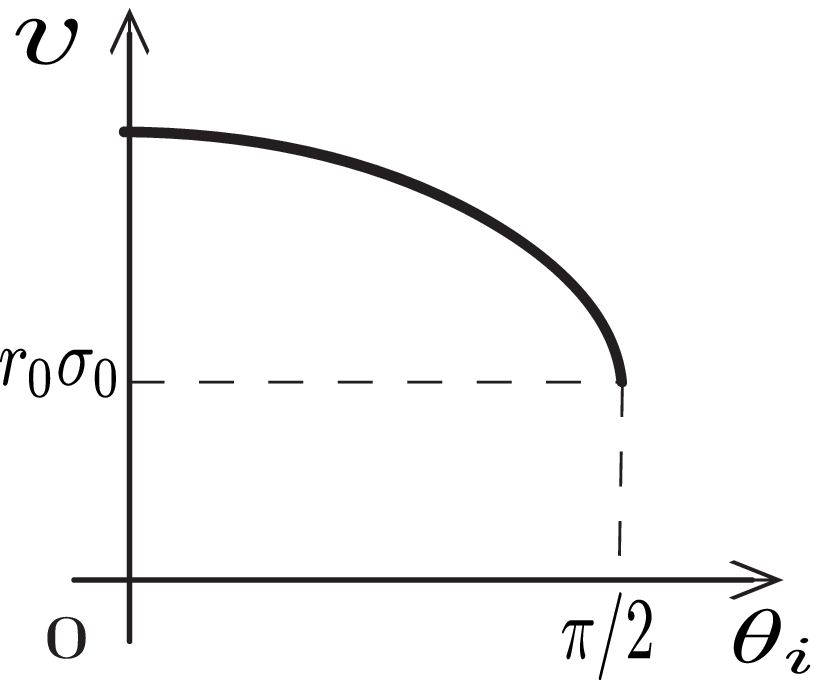}
  \end{center}
 \end{minipage} \\
\end{tabular}
\caption{Four types of $(n+1)$-surfaces on $vr$ and $v\theta_i$ planes.}
\end{figure}

\subsection{four types of $(n+1)$-surfaces and the condition for negative both null expansions}\label{sec:(n+1)-surfaces}
\subsubsection{Type A1 surface}

In region I, we introduce the $(n+1)$-surface of class A  
which is the topological disk given by the hyperboloid 
 	\begin{equation} \label{eq:hyperboloids}
 	 V=t_0+\rho-\sqrt{\rho^2+k^2},\quad R=\rho
	\end{equation}
where $t_0$ and $k$ are positive constants, 
and $t_0<k$.
We call this surface {\it type A1}.
In this surface, we find 
	\begin{equation}\label{eq:hyperboloids_derivative}
         V'=1-\frac{R}{\sqrt{R^2+k^2}},\quad 
         V''=-\frac{k^2}{(R^2+k^2)^{3/2}},\quad 
         R'=1\quad \mathrm{and}\quad R''=0. 
        \end{equation}
Substituting Eq.~(\ref{eq:hyperboloids_derivative}) and $m=0$
into Eq.~(\ref{eq:type1_0}), 
we obtain negative both null expansions as follows:
	\begin{equation}\label{eq:expansion_minkowski}
	 \vartheta_{(\pm)}=-\frac{(n+1)}{\sqrt{2}k}<0.
	\end{equation}
As noted in Sec.~\ref{sec:expansion_CIS}, 
both null expansions of class A are written in the same form as Eq.~(\ref{eq:expansion_minkowski}) for any $i$.
Thus, all type A1 surfaces have negative both null expansions.
\subsubsection{Type A2 surface}

In region II, we introduce the $(n+1)$-surface of class A 
in which $V$ and $R$ satisfy the condition 
 	\begin{equation}\label{eq:inverse_x}
	 \frac{dV}{dR}=\frac{a}{b-X^n},\quad R=\rho,
	\end{equation}
where $X=V/R$, $a$ and $b$ are positive constants. 
We call this surface {\it type A2}.
In this type, we find 
	\begin{equation}\label{eq:Vaidya_derivative}
	 V'=\frac{a}{b-X^n}, \quad 
         V''=\frac{naX^{n-1}}{R(b-X^n)^3}\left(X^{n+1}-bX+a\right),
         \quad R'=1\quad \mathrm{and}\quad R''=0.
	\end{equation}
Substituting Eq.~(\ref{eq:Vaidya_derivative}) and $m=\mu v^n$
into Eq.~(\ref{eq:type1_0}),
we obtain the following both null expansions for type A2:
	\begin{equation}\label{eq:vaidya_expansion}
	\vartheta_{(\pm)}=-N\left[\left(\mu a -n\right)\left\{naX^{n-1}+3n(b-a)X^n+\left(2\mu a-n\right)X^{2n}\right\}
        +n^2(2b-a)(b-a)\right],
	\end{equation}
where 
	\begin{equation}
	 N=\frac{\sqrt{n}}{R\sqrt{2a}}\left\{n(2b-a)+2X^n\left(\mu a-n\right)\right\}^{-3/2}.
	\end{equation}
To make both null expansions negative in Eq.~(\ref{eq:vaidya_expansion})
we impose the following sufficient conditions:
	\begin{eqnarray}
          b&>&a,\label{eq:condition_vaidya}\\
          \mu a&>&n.\label{eq:muaX}
        \end{eqnarray}
Thus, if conditions~(\ref{eq:condition_vaidya}) and~(\ref{eq:muaX}) are satisfied, 
both null expansions of type A2 are negative for any $i$.

\subsubsection{Type AB surface}

In region III, 
we introduce two types of $(n+1)$-surfaces. 
The first one is the type of class A which satisfies
	\begin{equation}
	 V=\rho,\quad R=r_0
	\end{equation}
where $r_0$ is the positive constant.
We call this surface {\it type AB}.
Both null expansions of type AB are 
	\begin{equation}\label{eq:mathcing00}
	 \vartheta_{(\pm)}=\frac{1}{\sqrt{2\Delta}}\frac{n}{R}
         \left(1-\frac{M}{nR^{n}}\right), 
	\end{equation}
where we have substituted $V'=1$, $V''=0$, $R'=R''=0$ and $m=M$
into Eq.~(\ref{eq:type1_0}).
Negative both null expansions are obtained, 
if and only if the condition $M>nR^{n}$ is satisfied.
For convenience, we define $\gamma=nR^n/M$. 
Then, if and only if $\gamma$ satisfies the condition 
	\begin{equation}\label{eq:gamma1}
	 0<\gamma<1,
	\end{equation}
both null expansions~(\ref{eq:mathcing00}) are negative. 

\subsubsection{Type B1 surface}

In region III, 
we have introduced two types of $(n+1)$-surfaces. 
The first one has been type AB.
The second is the type of class B 
which is a capping disk defined by 
 	\begin{equation}\label{eq:quadrant_of_circle}
	 \Theta_i^2+\left(\frac{\rho}{r_0}-\sigma_0\right)^2=\frac{\pi^2}{4}, \quad V=\rho
	\end{equation}
where $\sigma_0$ is the positive constant. 
We call this surface {\it type B1}.
Derivatives of type B1 are
	\begin{equation}\label{eq:derivative_in_schwarz}
         V'=1,\quad V''=0,\quad \Theta_i'=-\frac{\widetilde{V}}{r_0}\left(\frac{\pi^2}{4}-\widetilde{V}^2\right)^{-1/2}
          \quad\mathrm{and}\quad \Theta_i''=-\frac{\pi^2}{4r_0^2}\left(\frac{\pi^2}{4}-\widetilde{V}^2\right)^{-3/2},
        \end{equation}
where $\widetilde{V}=(V/r_0)-\sigma_0$.
As calculated in Sec.~\ref{sec:uniform_rad}, 
the expression of both null expansions of class B depends on $i$.
Substituting Eq.~(\ref{eq:derivative_in_schwarz}) and $m=M$ into Eq.~(\ref{typeB1}),
we obtain both null expansions of type B1 for the case of $i=1$ as follows:
        \begin{equation}\label{eq:schwarzschild_expansion01}
	 \vartheta_{1}{}_{(\pm)}{}=
         A_1\left[\pm\frac{\pi^2}{4}-\sqrt{\widetilde{\Delta}_1}
         \left\{n\left(\frac{1}{\gamma}-1\right)\left(\frac{\pi^2}{4}-\widetilde{V}^2\right)+(n+1)\widetilde{V}^2\right\}
         \pm n\widetilde{\Delta}_1\sqrt{\frac{\pi^2}{4}-\widetilde{V}^2}\cot\Theta_1\right]
        \end{equation}
where $\widetilde{\Delta}_1=\Sigma\left\{(\pi^2/4)-\widetilde{V}^2\right\}+\widetilde{V}^2$ and
$A_1=\sqrt{\Sigma}/(\sqrt{2}r_0\widetilde{\Delta}_1^{3/2})$. 
We obtain negative both null expansions, 
if the condition
	\begin{equation}\label{eq:typeB1_exp}
         \sqrt{\widetilde{\Delta}_1}
         \left\{n\left(\frac{1}{\gamma}-1\right)\left(\frac{\pi^2}{4}-\widetilde{V}^2\right)+(n+1)\widetilde{V}^2\right\}
         >
         \frac{\pi^2}{4}+n\widetilde{\Delta}_1\sqrt{\frac{\pi^2}{4}-\widetilde{V}^2}\cot\Theta_1
        \end{equation}
is satisfied. 
On the other hand, 
substituting Eq.~(\ref{eq:derivative_in_schwarz}) and $m=M$ into Eq.~(\ref{typeBall}),
we obtain both null expansions of type B1 for the case of $i\neq1$ as follows:
        \begin{eqnarray}\label{eq:schwarzschild_expansion02}
	 \nonumber \vartheta_{i}{}_{(\pm)}{}&=&
         A_i\left[\pm\frac{\pi^2}{4}\prod_{k=1}^{i-1}\sin\phi_k-\sqrt{\widetilde{\Delta}_i}
         \left\{n\left(\frac{1}{\gamma}-1\right)\left(\frac{\pi^2}{4}-\widetilde{V}^2\right)+(n+1)\widetilde{V}^2\prod_{k=1}^{i-1}\sin^2\phi_k\right\}\right.\\
         &&\left.\pm n\widetilde{\Delta}_i\sqrt{\frac{\pi^2}{4}-\widetilde{V}^2}\cot\Theta_i\prod_{k=1}^{i-1}\frac{1}{\sin\phi_k}\right]
        \end{eqnarray}
where $\widetilde{\Delta}_i=\Sigma\left(\pi^2/4-\widetilde{V}^2\right)+\widetilde{V}^2\prod_{k=1}^{i-1}\sin^2\phi_k$ and
$A_i=\sqrt{\Sigma}/(\sqrt{2}r_0\widetilde{\Delta}_i^{3/2})$. 
If the condition
	\begin{equation}\label{eq:typeB1_exp2}
         \sqrt{\widetilde{\Delta}_i}
         \left\{n\left(\frac{1}{\gamma}-1\right)\left(\frac{\pi^2}{4}-\widetilde{V}^2\right)+(n+1)\widetilde{V}^2\prod_{k=1}^{i-1}\sin^2\phi_k\right\}
         >
         \frac{\pi^2}{4}\prod_{k=1}^{i-1}\sin\phi_k+n\widetilde{\Delta}_i\sqrt{\frac{\pi^2}{4}-\widetilde{V}^2}\cot\Theta_i\prod_{k=1}^{i-1}\frac{1}{\sin\phi_k}
        \end{equation}
is satisfied, 
we obtain negative both null expansions. 
However, when we take the limit $\phi_k$ to $0$, the second term on the right-hand side of Eq.~(\ref{eq:typeB1_exp2}) has 
infinitely large value. 
From this reason, we cannot obtain negative both null expansions of 
type B1 with $\phi_k=0$ in the case of $i\neq1$.
To avoid this difficulty from now on we only focus on type B1 with $i=1$.

\subsection{Matching of four types of $(n+1)$-surfaces}\label{sec:matching}

We shall match four types of $(n+1)$-surfaces given in the previous section. 
Schematic figures of these surfaces are shown in Fig.~3. 
	\begin{figure}[h!]
        \begin{center}
        \subfigure[]{
   	\includegraphics[width=.35 \columnwidth]{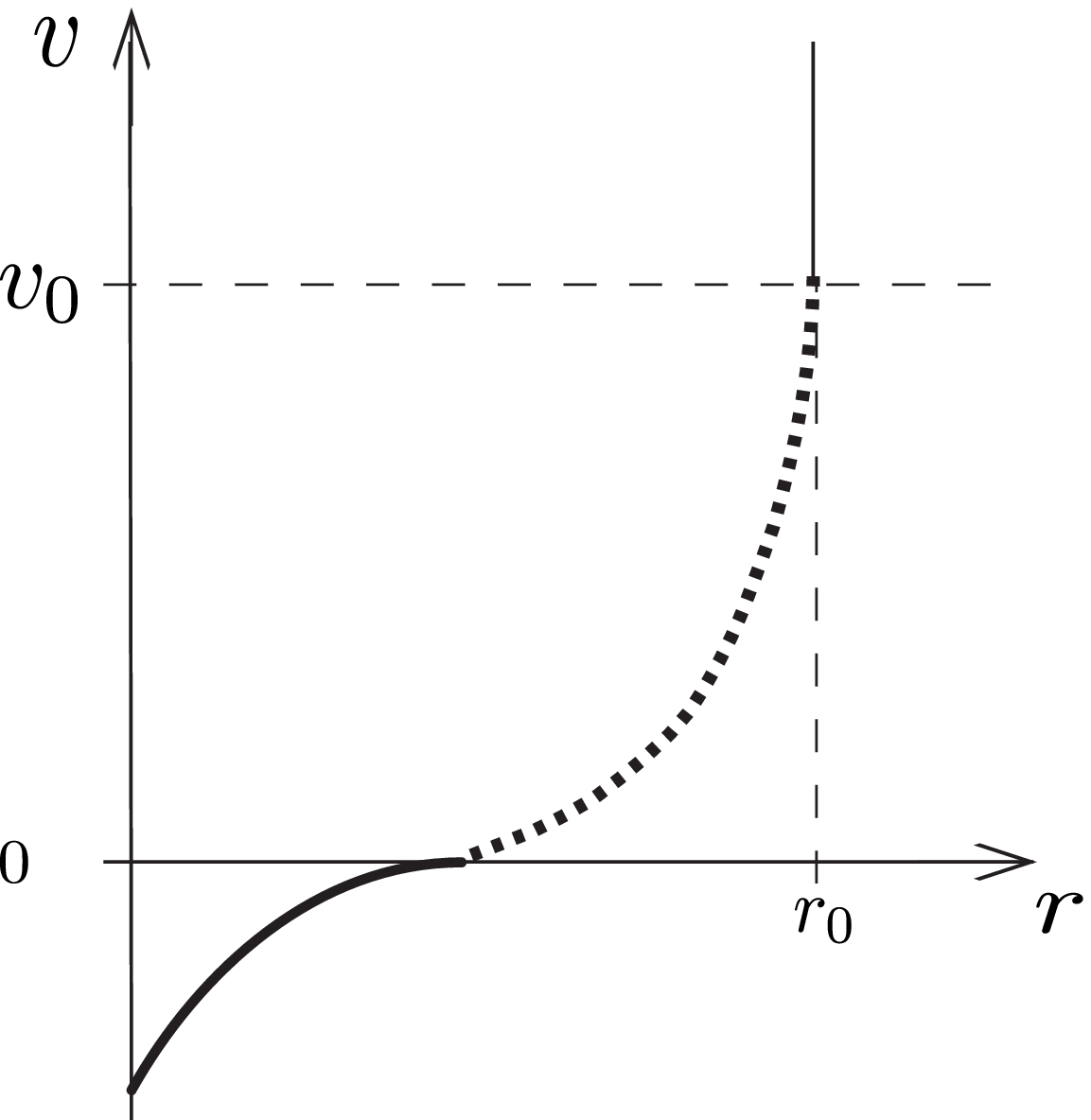}
  	\label{fig:circle}}~\quad\quad
  	\subfigure[]{
   	\includegraphics[width=.40 \columnwidth]{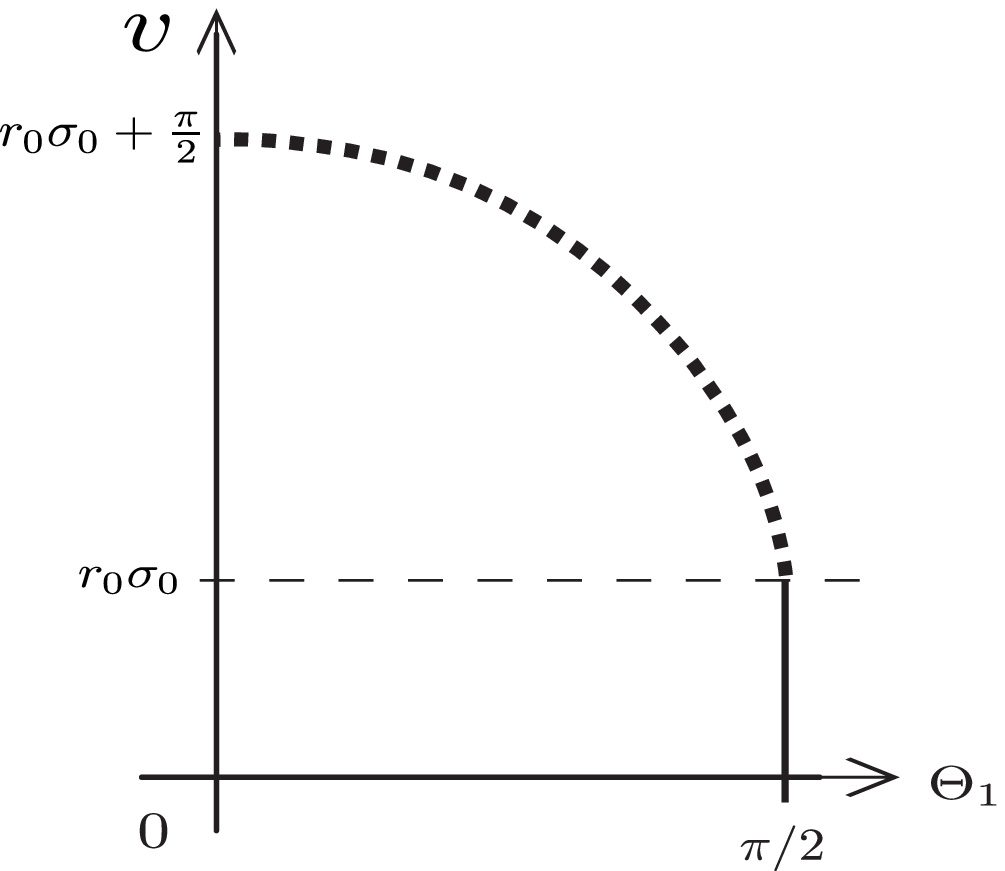}
 	\label{fig:matching_surfaces}}
   	      \caption{Schematic figures of the matching surface.
              (a) Schematic figure of
              type A1, A2, and AB surfaces on the $vr$ plane, where
              the thick solid line, the dotted line and the solid line 
              are type AI, type A2 and 
              type AB, respectively.
              (b) Schematic figure of type AB and type B1 on the $v\Theta_1$ plane, 
              where the solid line and the dotted line are type AB and type B1, respectively.
              }
        \end{center}
	  \end{figure}
We match a type A1 surface with a type A2 surface on $v=0$, 
the type A2 surface with a type AB surface on $v=v_0$, 
and the type AB surface with a type B1 surface on $v=r_0\sigma_0$ in region III.

\subsubsection{Matching type A1 with type A2}

We match the type A1 surface with the type A2 surface on $v=0$.
In order to obtain a smooth matching surface we match the derivative of both surfaces.
From Eq.~(\ref{eq:hyperboloids_derivative}) we know that the derivative $dv/dr$ of 
type A1 is 
positive and is less than one
	\begin{equation}\label{eq:derivative_cond00}
	 0<\frac{dv}{dr}<1.
	\end{equation}
Thus, the derivative of type A2 must 
satisfy this condition (\ref{eq:derivative_cond00}) on $v=0$.
Substituting $v=0$, i.e., $X=0$ into Eq.~(\ref{eq:inverse_x}), 
and imposing the condition (\ref{eq:derivative_cond00}) on it, 
we obtain the condition
	\begin{equation}\label{eq:ab}
	 0<\frac{a}{b}<1. 
	\end{equation}
Note that this condition is written in the same form 
as the condition~(\ref{eq:condition_vaidya}).

\subsubsection{Matching type A2 with type AB}

We match the type A2 surface with the type AB surface on $v=v_0$.
Note that the derivative $dv/dr$ of type AB is infinitely large 
	\begin{equation}\label{condition_infty}
	 \frac{dv}{dr}\rightarrow \infty.
	\end{equation}
Choosing $b=X^n$ in Eq.~(\ref{eq:inverse_x}), 
we can match the derivative of the type AB surface with that of type A2.
Thus, in region II, $X$ satisfies $0<X^n<b$.
On $v=v_0$ the type AB surface has the constant radius $r=r_0$.
Substituting $r=r_0$ and $v=v_0$ into $b=X^n=(v/r)^n$, 
we express $b$ as
	\begin{equation}\label{eq:b}
	 b=\frac{n}{\mu\gamma}, 
	\end{equation}
where $\gamma=nr_0^n/M$. 

It should be noted that in the study of Bengtsson and Senovilla, 
they rewrote the derivative $dV/dR$ as follows
	\begin{equation}\label{eq:rewrite}
	 R\frac{dX}{dR}=\frac{F(X)}{b-X}
	\end{equation}
where $F(X)=X^2-bX+a$, 
and demanded $F(X)>0$ to obtain the solution they wanted.
In our study
Eq.~(\ref{eq:rewrite}) becomes 
	\begin{equation}\label{eq:rewrite2}
	 R\frac{dX}{dR}=\frac{G(X)}{b-X^n}
	\end{equation}
where $G(X)=X^{n+1}-bX+a$.
Similarly to Bengtsson and Senovilla's study, 
we shall impose $G(X)>0$.
The minimum value of $G(X)$ is given by 
	\begin{equation}\label{eq:minimum}
	 G(X_{\mathrm{min}})=a-\frac{n}{(n+1)^{(n+1)/n}}b^{(n+1)/n},
	\end{equation}
where $X_{\mathrm{min}}=(b/(n+1))^{1/n}$, since
	\begin{equation}
	 \frac{dG(X_{\mathrm{min}})}{dX}=(n+1)X_{\mathrm{min}}^{n}-b=0.
	\end{equation}
To get $G(X)>0$, we impose the condition on parameters $a$ and $b$ 
from the following two cases:
(a) $X_{\mathrm{min}}<b$ and $G(X_{\mathrm{min}})>0$, (b) $b<X_{\mathrm{min}}$ and $G(b)>0$.
Note that in the four-dimensional ($n=1$) spacetime we get $X_{\mathrm{min}}=b/2$.
Thus, we cannot use case (b) in this spacetime.
Substituting $n=1$ into $G(X_{\mathrm{min}})>0$, we obtain the condition as
	\begin{equation}\label{eq:(a)_4dim}
	 a>\frac{b^2}{4}.
	\end{equation}
This condition is given in Bengtsson and Senovilla's study~\cite{I.Bengtsson}.
We show the allowed region of conditions~(\ref{eq:ab}) and (\ref{eq:(a)_4dim}) on the $ab$ plane
in Fig.~4(a).
On the other hand, we shall consider the $(n+3)$-dimensional spacetime with $n>1$.
In case (a), substituting $X_{\mathrm{min}}=(b/(n+1))^{1/n}$ into conditions $X_{\mathrm{min}}<b$ and 
$G(X_{\mathrm{min}})>0$, 
and after some calculation, 
we find 
	\begin{equation}\label{eq:(a)}
	 b>K,\quad\mathrm{and}\quad a>n\left(\frac{b}{n+1}\right)^{(n+1)/n},
        \end{equation}
where $K=(n+1)^{1/(1-n)}$.
In case (b), we get 
	\begin{equation}\label{eq:(b)}
	 b<K \quad
	 \mathrm{and}\quad a>b^{2}-b^{n+1},
        \end{equation}
where we have rewritten conditions $b<X_{\mathrm{min}}$ and 
$G(b)>0$.
We also show the allowed region of conditions on parameters $a$ and $b$ as in Fig.~5(a).
The allowed region of conditions~(\ref{eq:ab}) and~(\ref{eq:(a)}) 
is shown by the dark-shaded area on the $ab$ plane, 
while the pale-shaded area shows 
the allowed region of conditions~(\ref{eq:ab}) and
~(\ref{eq:(b)}).
Note that in Fig.~5(a) we have substituted $n=2$ into these conditions for reference. 

	\begin{figure}[h!]
        \begin{center}
        \subfigure[]{
   	\includegraphics[width=.4 \columnwidth]{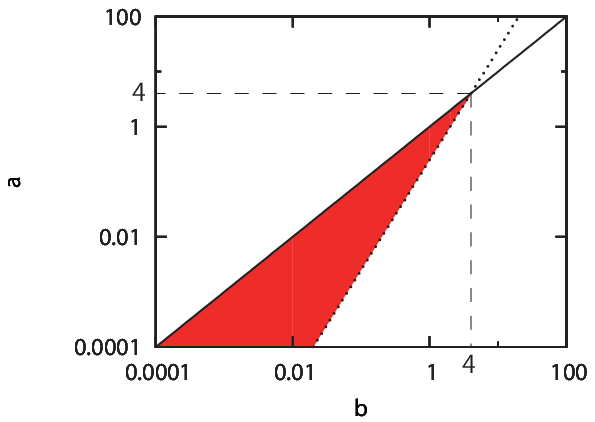}
  	}~\quad\quad
  	\subfigure[]{
   	\includegraphics[width=.4 \columnwidth]{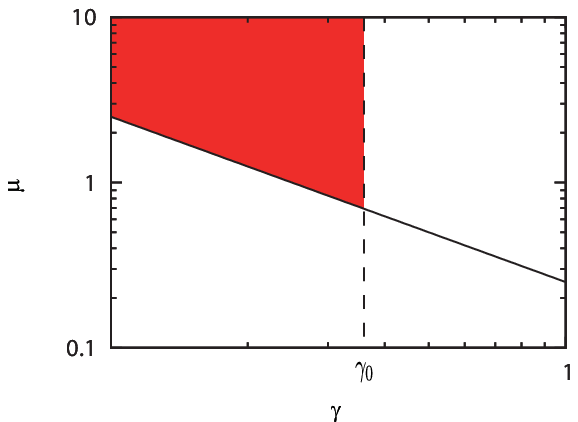}
 	}
         \caption{
              (a)Schematic figure of the allowed region for conditions on the $ab$ plane.
              The solid line and the dotted line 
              are $a=b$ and $a=b^2/4$, respectively.
              The shaded area is the allowed region of conditions~(\ref{eq:ab}) and (\ref{eq:(a)_4dim}).
              (b)Schematic figure of the allowed region for conditions on the $\mu\gamma$ plane.
              The solid line denotes $\mu=1/(4\gamma)$.
              $\gamma_0$ is the upper bound of $\gamma$ imposed by the condition (\ref{eq:condition01}).
              The allowed region of conditions~(\ref{eq:condition01}) and (\ref{eq:mu_4dim})
              is shown by the shaded area. 
              }
        \end{center}
	  \end{figure}

\subsubsection{Matching type AB with type B1}

We match the type AB surface with the type B1 surface on $v=r_0\sigma_0$ in region III.
From Fig.~3(b) 
the derivative $d\theta/dv$ of these surfaces vanishes on $v=r_0\sigma_0$.
Thus, we can smoothly match these surfaces on $v=r_0\sigma_0$.
On $v=r_0\sigma_0$, i.e., $\widetilde{V}=0$ the condition~(\ref{eq:typeB1_exp}) 
becomes 
	\begin{equation}\label{eq:condition01}
	 n\sqrt{\frac{2}{\gamma}-1}\left(\frac{1}{\gamma}-1\right)>\frac{2}{\pi}
	\end{equation}	
where we have substituted $\widetilde{V}=0$ into Eq.~(\ref{eq:typeB1_exp}). 
Now, we shall find the upper bound on $\gamma$ in Eq.~(\ref{eq:condition01}).
The solution of Eq.~(\ref{eq:condition01}) is
	\begin{equation}\label{eq:gamma_Gamma}
	 \gamma<\gamma_0=\frac{6}{6+\Gamma},
	\end{equation}
where 
$\gamma_0$ is the upper bound on $\gamma$, 
	\begin{equation}\label{eq:Gamma00}
	 \Gamma=-1+\left(E+F\right)^{1/3}+\left(E+F\right)^{-1/3},
	\end{equation}
and 
	\begin{equation}
	 E=-1+54\left(\frac{2}{n\pi}\right)^2,\quad F=6\sqrt{-3\left(\frac{2}{n\pi}\right)^2+
         81\left(\frac{2}{n\pi}\right)^4}.
	\end{equation}
When we choose $n=1$, $\gamma_0\simeq 0.6851$ which coincides with 
the upper bound on $\gamma$ derived in Ref.~\cite{I.Bengtsson}.
For $n>3$, $F$ becomes a pure imaginary number.
Then we can express $\Gamma$ as
	\begin{equation}
	 \Gamma=-1+2\cos\left(\frac{\zeta}{3}\right),
	\end{equation}
where $\zeta=\arccos(E)$.
$\zeta$ is the monotonically decreasing function of $n$.
In the large $n$ limit $\zeta=\pi$, while 
for $n=4$, $\zeta\simeq1.892$.
Thus, $\gamma_0$ monotonically increases and 
approaches one in the large $n$ limit.
On the other hand, $\gamma_0$ gets a minimum value 
$\gamma_{\mathrm{min}}=6/(6+\Gamma_2)\simeq0.7946$ on $n=2$,
where $\Gamma_2=-1+(E_2+F_2)^{1/3}+(E_2+F_2)^{-1/3}$,
$E_2=-1+(54/\pi^2)$ and $F_2=6\sqrt{-3+(81/\pi^2)}/\pi$.
Note that the condition~(\ref{eq:condition01}) on $\gamma$ is stronger
than the condition~(\ref{eq:typeB1_exp}) on it.

Now we shall discuss the physical meaning of $\gamma_0$.
For this purpose, we introduce the following variable:
	\begin{equation}
	 \bar{\gamma}_0=\left(\frac{\gamma_0}{2}\right)^{1/n}=\frac{r_0}{r_g}
	\end{equation}
where $r_g=(2M/n)^{1/n}$ is the Schwarzschild-Tangherlini radius,
and $r_0$ is the maximum radius of the trapped surface considered here.
Thus, $\bar{\gamma}_0$ denotes the maximum radius of 
the trapped surface normalized by $r_g$ for each dimension. 
We can find that $\bar{\gamma}_0$ also monotonically increases and 
also approaches one in the large $n$ limit.
Therefore, in the large $n$ limit, the maximum radius of the trapped surface 
approaches $r_g$.
Note that when we take the large $n$ limit 
the $vv$-component in the $(n+3)$-dimensional Vaidya metric~(\ref{eq:metric})
does not approach zero. 
Instead, it approaches one in this limit.

\subsection{Condition on the mass parameter}\label{sec:condition}

In previous sections,
we have imposed several conditions on parameters to obtain a trapped surface.
We shall combine these conditions and shall get the condition on the mass parameter $\mu$.

In the four-dimensional case, 
from Fig.~4(a) we can see 
$0<a<4$ and $0<b<4$, respectively.
Substituting Eq.~(\ref{eq:b}) with $n=1$ into $b<4$, we get the condition on $\mu$ as 
	\begin{equation}\label{eq:mu_4dim}
	 \mu>\frac{1}{4\gamma}.
	\end{equation}
In Fig.~4(b), 
we show the allowed region of conditions (\ref{eq:condition01}) and (\ref{eq:mu_4dim}).
From this figure we can understand that 
if the mass function satisfies $\mu>1/(4\gamma_0)\simeq0.3649$, 
we can obtain a trapped surface extended into region I.
This result is given in Ref.~\cite{I.Bengtsson}.

Next, we shall discuss the $(n+3)$-dimensional spacetime
with $n>1$.
From the dark-shaded area in Fig.~5(a) we can see that
$a$ and $b$ satisfy $L<a<N$ and $K<b<N$, respectively, 
where $K=(n+1)^{1/(1-n)}$, $L=n(n+1)^{(n+1)/(1-n)}$ and $N=(n+1)^{n+1}/n^n$.
Combining $a<N$ with Eq.~(\ref{eq:muaX}), we get the condition on $\mu$ as 
	\begin{equation}\label{eq:mu(a)1}
	 \mu>\frac{n}{N}.
	\end{equation}
Substituting Eq.~(\ref{eq:b}) into $K<b<N$, 
we also get the condition on $\mu$ as 
	\begin{equation}\label{eq:mu(a)2}
	 \frac{n}{K\gamma}>\mu>\frac{n}{N\gamma}.
	\end{equation}
We combine conditions~(\ref{eq:condition01}), (\ref{eq:mu(a)1}) and (\ref{eq:mu(a)2}), 
and show the allowed region of this combined condition in Fig.~5(b).
In Fig.~5(b), the pale-shaded area is the allowed region of this combined condition, 
where we have substituted $n=2$ into these conditions for reference.
On the other hand, 
from the pale-shaded area in Fig.~5(a)
we can see that $a$ and $b$ are bounded above by $K$.
Combining $a<K$ with Eq.~(\ref{eq:muaX}), we get the condition on $\mu$ as 
	\begin{equation}\label{eq:mu(b)1}
	 \mu>\frac{n}{K}.
	\end{equation}
Substituting Eq.~(\ref{eq:b}) into $b<K$, 
we also get the condition on $\mu$ as 
	\begin{equation}\label{eq:mu(b)2}
	 \mu>\frac{n}{K\gamma}.
	\end{equation}
We also combine conditions~(\ref{eq:condition01}), (\ref{eq:mu(b)1}) and (\ref{eq:mu(b)2}), 
and show the allowed region of this combined condition 
by the dark-shaded area in Fig.~5(b).
	\begin{figure}[h!]
         \begin{center}
         \subfigure[]{
   	 \includegraphics[width=.4 \columnwidth]{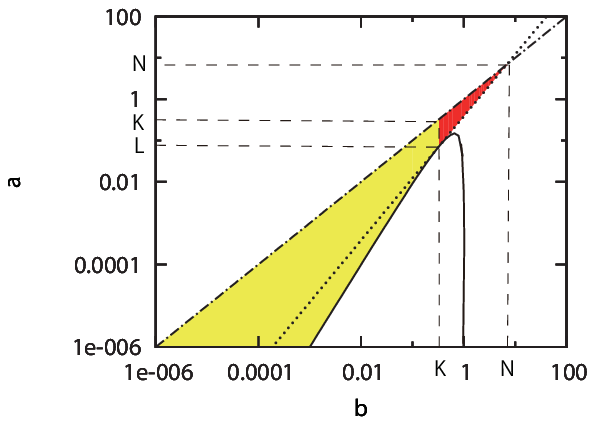}
  	 }~\quad\quad
  	 \subfigure[]{
   	 \includegraphics[width=.4 \columnwidth]{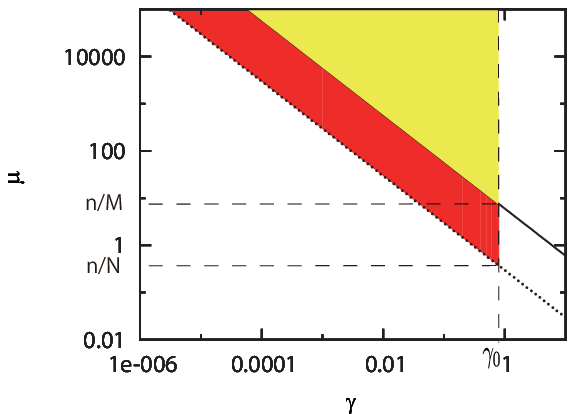}
 	 }
         \caption{
              (a)Schematic figure of the allowed region for conditions on the $ab$ plane.
              the dotted-dashed line, the dotted line and the solid line 
              are $a=b$, $a=n({b}/{n+1})^{(n+1)/n}$ and $a=b^2-b^{n+1}$, respectively.
              The dark-shaded area 
              shows the allowed region of 
              conditions~(\ref{eq:ab}) and 
              (\ref{eq:(a)}), while 
              the pale-shaded area shows 
              the allowed region 
              of conditions~(\ref{eq:ab}) and 
              (\ref{eq:(b)}),
              where $K=(n+1)^{1/(1-n)}$, 
              $L=n(n+1)^{(n+1)/(1-n)}$ and $N=(n+1)^{n+1}/n^n$.
              Note that in this figure 
              we have substituted $n=2$ into these conditions for reference. 
              (b)Schematic figure of the allowed region for conditions on $\mu\gamma$ plane,
              where the solid line and the dotted line are $\mu=n/(K\gamma)$ and
              $\mu=n/(N\gamma)$, respectively.
              The allowed region of conditions~(\ref{eq:condition01}), (\ref{eq:mu(a)1}) and (\ref{eq:mu(a)2})
              is shown by the dark-shaded area, 
	      while the pale-shaded area shows 
	      the allowed region of conditions~(\ref{eq:condition01}), (\ref{eq:mu(b)1}) and (\ref{eq:mu(b)2}).
              $\gamma_0$ is the upper bound of $\gamma$ imposed by the condition (\ref{eq:condition01}).
              Note that in this figure 
              we have substituted $n=2$ into these conditions for reference.
              }
        \end{center}
	  \end{figure}
Fig.~5(b) shows that $\mu$ is bounded below by $\mu=n/N$.
Thus, at least if $\mu>n/(N\gamma)$ and $\gamma<\gamma_0$, 
we can construct a trapped surface extended into region I
by the appropriate choice of $a$ and $b$.
$n/N$ is a monotonically increasing function 
and it approaches $1/e$ in the large $n$ limit. 
On the other hand, $\gamma_0$ is also the monotonically increasing function. 
$\gamma_0$ approaches one in the large $n$ limit, 
while it has the minimum value $\gamma_{\mathrm{min}}$ for $n=2$.
Thus, $1/\gamma_0$ gets the maximum value $1/\gamma_{\mathrm{min}}\simeq 1.258$ on $n=2$.
To put all the conditions together,
we conclude that 
in the $(n+3)$-dimensional self-similar Vaidya spacetime spacetime with $n>1$ 
if the mass parameter $\mu$ is greater than $1/(\gamma_{\mathrm{min}}e)\simeq0.4628$, 
we can construct a trapped surface extended into region I.

\subsection{Naked singularity and the trapped surface}\label{sec:singularity}

In Bengtsson and Senovilla's study, they investigated the four-dimensional self-similar Vaidya spacetime and 
showed that there is no naked singularity,
if the spacetime has a trapped surface extended into region I~\cite{I.Bengtsson}.
How about this feature in the higher-dimensional spacetime?
From this context, we shall discuss naked singularity in the $(n+3)$-dimensional self-similar Vaidya spacetime with $n>1$. 
In Sec.~\ref{sec:vaidya} we have mentioned that if and only if the mass parameter satisfies the condition~(\ref{def:ns}), 
there exists naked singularity.
On the other hand, in Sec.~\ref{sec:condition} 
we have shown the condition~$\mu>1/(\gamma_{\mathrm{min}}e)$ on the mass parameter 
so as to construct a trapped surface extended into region I. 
Comparing both conditions, 
we find 
	\begin{equation}
	 \left(\frac{1}{2}\right)^{n+1}<\frac{1}{\gamma_{\mathrm{min}}},\quad \mathrm{and}\quad
         \left(\frac{n}{n+1}\right)^{n+1}<\frac{1}{e}.
	\end{equation}
Thus, both conditions on the mass parameter are inconsistent with each other.
Therefore, we conclude that 
if a trapped surface extended into region I can be constructed by the above discussion, 
there is no naked singularity in the $(n+3)$-dimensional self-similar Vaidya spacetime.

\section{Conclusion and discussions}\label{sec:conclusion}

We have investigated a trapped surface and naked singularity 
in the $(n+3)$-dimensional self-similar Vaidya spacetime.
A trapped surface is defined as the closed spacelike $(n+1)$-surface which has negative both null expansions.
To construct a trapped surface
we have introduced two classes of $(n+1)$-surfaces 
and have made both null expansions of these surfaces negative.
We also have introduced four types of $(n+1)$-surfaces and matched smoothly 
the type A1 surface with the type A2 surface on $v=0$, 
the type A2 surface with the type AB surface on $v=v_0$, 
and the type AB surface with the type B1 surface on $v=r_0\sigma_0$.
To obtain the smooth matching and negative both null expansions 
we have imposed conditions on parameters.
Putting these conditions together, we have got 
the condition on the mass parameter, i.e.,
we have got the lower limit $1/(\gamma_{\mathrm{min}}e)\simeq0.4628$ on the mass parameter
for $n>1$.
Therefore, we have shown that 
in the $(n+3)$-dimensional self-similar Vaidya spacetime for $n>1$,
if the mass parameter is greater than $1/(\gamma_{\mathrm{min}}e)\simeq0.4628$, 
we get a trapped surface extended into region I.
Moreover, we have found that the maximum radius of the trapped surface constructed in this study 
monotonically increases for $n$.
If we take the limit $n$ to infinity, the maximum radius of the trapped surface
approaches the Schwarzschild-Tangherlini radius.
Also, we have shown that 
there is no naked singularity, 
if the trapped surface given by the above construction exists
in the $(n+3)$-dimensional self-similar Vaidya spacetime. 
These results are similar to those of Bengtsson and Senovilla in the four-dimensional case. 
However, the conditions are affected by the spacetime dimension, 
which can be seen in Figs.~4 and 5.

The trapped surface constructed in this study satisfies 
the condition $\gamma<1$.
This surface exists inside the black hole, 
because the event horizon satisfies the condition $\gamma=2$.
Although in the higher-dimensional spacetime 
we can have the large variety of $(n+1)$-surfaces, 
while we have introduced two classes of $(n+1)$-surfaces in this paper.
Thus, if we introduce other kinds of $(n+1)$-surfaces and choose 
the appropriate matching for these surfaces, 
we can construct other kinds of trapped surfaces 
extended into region I.

Although Eardley's conjecture is considered only in four-dimensional spacetimes, 
applying this conjecture into the general dimension might be fruitful to 
define dynamical black holes in asymptotically flat higher-dimensional spacetimes.

\section*{ACKNOWLEDGMENTS}

We are very grateful to H.~Ohmiya for fruitful discussion.
We would like to thank J.~M.~M.~Senovilla, A.~Ishibashi, D.~Ida, K.~Nakao and T.~Shiromizu for helpful comment. 
TH was partly supported by the Grant-in-Aid for Scientific Research
Fund of the Ministry of Education, Culture, Sports, Science and Technology, Japan [Young Scientists (B) 
21740190
].


\end{document}